%% file: MainJuly17.tex
\documentclass[12pt]{article}
\usepackage[margin = 1.25in]{geometry}
\usepackage[utf8]{inputenc}
\usepackage{amsmath, tabularx, mathtools ,soul}
\usepackage{enumitem}
\usepackage{amsfonts}
\usepackage{comment}
\usepackage{multicol}
\usepackage{float}
\usepackage{cleveref}
\usepackage{bm}
\usepackage{graphicx,xcolor}
\usepackage{multicol}
\usepackage{bbm}
\usepackage{changepage}
\usepackage{tikz}
\usepackage{todonotes}
\title{Information Acquisition \\and Time-Risk Preference}
\author{Daniel Chen\footnote{dtchen@princeton.edu, Department of Economics, Princeton University} and Weijie Zhong\footnote{wjzhong@stanford.edu, Graduate School of Business, Stanford University \\ We thank Kayla Jiang for excellent research assistance.}}
\date{\today}

\usepackage{amsthm}

\newtheorem{theorem}{Theorem}
\newtheorem{lemma}{Lemma}
\newtheorem{corollary}{Corollary}[theorem]

\crefname{conj}{Conjecture}{Conjectures}

\theoremstyle{definition}
\newtheorem{definition}{Definition}
\allowdisplaybreaks
\usepackage{natbib}
\usepackage{setspace}
\begin{document}

\maketitle
\onehalfspacing

\begin{abstract}
An agent acquires information dynamically until her belief about a binary state reaches an upper or lower threshold. She can choose any signal process subject to a constraint on the rate of entropy reduction. Strategies are ordered by ``time risk"---the dispersion of the distribution of threshold-hitting times. 
We construct a strategy maximizing time risk (\emph{Greedy Exploitation}) and one minimizing it (\emph{Pure Accumulation}). Under either strategy, beliefs follow a compensated Poisson process. In the former, beliefs jump to the threshold that is closer in Bregman divergence. In the latter, beliefs jump to the unique point with the same entropy as the current belief. 
\end{abstract}

\section{Introduction}

In this paper, we study information acquisition by a Bayesian agent about an unknown binary state that may be either zero or one. The agent wants to be reasonably certain about the state and is satisfied once her posterior belief that the state is one reaches either an upper or a lower threshold. She earns a unit payoff when this happens. She has great flexibility in how she can learn but has a resource constraint that limits her rate of learning. That is, she can choose any posterior belief process subject to a constraint on the rate of entropy reduction.

Our simple model captures three important features of many economic settings: flexible learning, limited resources, and threshold decision rules.
These features often appear in the contexts of research and development, clinical trials, digital marketing, user-experience testing, and others.
For example, consider a data scientist at Facebook who must assess whether to introduce a new feature to the platform. The unknown state is whether adding the feature increases or decreases user engagement (and thus profits). The data scientist can learn about the state by conducting A/B tests. To provide incentives, her manager offers her a bonus when she has learned sufficiently precisely about the state (i.e. her A/B tests identify the state at some minimal level of statistical power\footnote{In our binary state model, there is a one-to-one mapping between thresholds and the statistical power of a test (that is, the likelihood of Type II errors).}). She can freely design many aspects of the A/B tests---e.g., she can select the subpopulation of users that she gives access to the feature and she can adjust the length of time a user has access to the feature among other choices. However, there are limits to what she can do that constrain her speed of learning---e.g., her manager does not allow her to implement the feature for all users simultaneously as doing so could be disastrous for profits if the feature is disliked or she can only implement the feature for a given user for a short amount of time, etc.


Our main contribution is to show how, in such settings, the agent's optimal learning strategy depends on her \emph{time-risk preferences}. That is, we allow for a rich set of preferences over threshold-hitting times beyond the standard case of exponential discounting. We say that the agent is time-risk loving (averse) whenever her utility over threshold-hitting times is convex (concave).\footnote{\label{fn:1}In this paper, we model the agent as an expected-utility maximizer. However, it is easy to see that all of our results will go through as long as the agent has a preference relation over stopping times that is monotonic in the mean-preserving spread order.} We derive a learning strategy that is optimal whenever the agent is time-risk loving and a strategy that is optimal whenever she is time-risk averse. Critically, the optimality of these strategies does not depend on the shape of the agent's utility function beyond its convexity or concavity.

In reality, there are many reasons why individuals may have time preferences that differ from the predominantly studied case of exponential discounting. This may be due to external factors such as explicit discounting, exogenous decision deadlines, and flow costs associated with foregone opportunities while learning. It may also be due to internal factors such as present bias resulting from hyperbolic discounting. We provide a simple framework that allows for these factors when studying optimal learning and derive strategies that are uniformly optimal up to the convexity or concavity of the utility function over threshold-hitting times.


We now briefly describe the two learning strategies and provide intuition for them. When the agent is time-risk loving, a \emph{Greedy Exploitation} strategy is optimal. Under this strategy, the agent myopically maximizes the instantaneous rate that her beliefs jump to a threshold. She acquires a rare but decisive signal that, upon arrival, induces her belief to jump to the threshold that is closest in the Bregman divergence.
By targeting the closer threshold, she can jump at a faster rate without violating her constraint on the rate of entropy reduction. In the absence of a signal arrival, her belief experiences compensating drift in the direction of the farther threshold. Eventually, her belief reaches a point that is equidistant in the Bregman divergence to the two thresholds. At this point, she acquires signals such that her beliefs may jump to either threshold but at rates set so that there is no net compensating drift so that her belief is stationary in the absence of a jump.

Intuitively, Greedy Exploitation is optimal because it produces a very risky distribution of threshold-hitting times. Because the strategy is greedy, it yields a high probability of an early hitting time. However, in the absence of a jump, since beliefs drift towards the farther threshold, the jump rate decreases so that the expected amount of time remaining until a threshold is reached \emph{increases}. In this sense, the agent makes no ``progress" in the absence of a jump. Thus there is a high probability of late threshold hitting times as well. We, in fact, show that among all strategies that exhaust the agent's resources (in that the constraint on the rate of entropy reduction is binding at all points in time), Greedy Exploitation yields a distribution of hitting times that is maximal in the mean-preserving spread order.  In this sense, it maximizes time risk.

When the agent is time-risk averse, she instead seeks to minimize time risk. In this case, an optimal strategy is \emph{Pure Accumulation}. Under this strategy, her beliefs reach a threshold at a \emph{deterministic} time. Her beliefs follow a compensated Poisson process that jumps in the direction of the threshold that is farther away but to an interior belief that has the same entropy as her current belief. In the absence of a jump, her belief experiences a compensating drift toward the closer threshold. Pure accumulation is a continuous-time analog of the ``suspense-maximal'' policy in \cite{ely2015suspense}. The strategy is, in effect, the ``opposite" of Greedy Exploitation. Because jumps are always to beliefs with the same entropy as the current belief, in the event of a jump, there is no progress: a jump does not reduce the expected amount of time until a threshold is hit. Instead, all progress is made through drift, which is why the threshold hitting time is deterministic. Thus, Pure Accumulation entails \emph{no} time risk. It, therefore, produces a distribution of hitting times that is minimal in the mean-preserving spread order among all strategies that exhaust the agent's resources (in that the constraint on the rate of entropy reduction is binding at all points in time).

Our analysis of optimal learning through the lens of time-risk preferences has implications for both information acquisition in practice and economic modeling. Our model predicts that an agent who is time-risk-loving should use Greedy Exploitation, whereas an agent who is time-risk-averse should opt for Pure Accumulation, provided the agent has access to these learning strategies. Moreover, whenever the agent's space of available learning strategies is sufficiently rich, Brownian signals are generally suboptimal. Thus, when writing models where agents acquire information with parameterized signal structures, economists should be mindful of whether these signal structures are without loss of optimality given the time-risk preferences of the agents they seek to model.



\section{Related Literature}\label{related literature}
Our paper contributes to a large literature on information acquisition. As in \cite{wald1947foundations} and \cite{arrow1949bayes} we study a sequential sampling problem but allow the agent to flexibly design the signal process as in \cite{zhong2017dynamic}, \cite{hebert2016rational}, \cite{hebert2021}, \cite{steiner2016rational}, and \cite{georgiadis2023information}. Whereas most of these papers restrict attention to the standard case of exponential discounting or a linear delay cost, we allow for more general time preferences.
For example, \cite{zhong2017dynamic} assumes exponential discounting which implies time-risk loving preferences whereas \cite{hebert2016rational} and \cite{hebert2021} assume a linear delay cost which implies time-risk neutral preferences.  Our results suggest that the assumed time-risk preferences dictate the qualitative features of the optimal strategies identified in these  papers.\footnote{\cite{hebert2021} allow for both discounting and a linear delay cost. However, they consider the time-risk neutral limit for the majority of their analysis. Their main objective is to study how different costs or constraints on information acquistion affect optimal learning strategies which is orthogonal to the objective of our paper. \cite{zhong2017dynamic} assumes a belief-dependent payoff function that does not have a threshold structure but shows that the optimal learning strategy is nevertheless similar to Greedy Exploitation.}

Pure Accumulation is a continuous-time variant of the suspense-maximal strategy introduced by \cite{ely2015suspense} in a discrete-time setting with a deterministic and finite time horizon. In \cite{ely2015suspense}, the suspense-maximal strategy is optimal in that it maximizes an expected utility that is increasing in the variances of future beliefs.
\cite{georgiadis2023information} also finds that a strategy similar to Pure Accumulation is optimal in a setting where the stopping time is \textit{exogenous} and does not depend on the learning strategy.\footnote{In our paper, the stopping is determined by the exogenous thresholds as well as the \textit{endogenously} chosen learning strategy.} The mechanisms behind the optimality of Pure Accumulation in both of these papers are distinct from that of this paper where it is shown to be time-risk aversion.


In our analysis, the key summary statistic that determines the payoff from a strategy is the distribution of the time that the agent's belief first reaches a threshold. This statistic defines a \emph{time lottery}, which is an object studied in an emerging literature on time-risk preferences. \cite{chesson2003commonalities} and \cite{chen2013effect} show that the expected discounted utility framework implies preferences that are \emph{risk seeking over time lotteries} (RSTL). \cite{dejarnette2020time} show that within a broad class of models RSTL can not be violated if there is stochastic impatience. However, experimental evidence suggests that subjects are often \emph{risk averse over time lotteries} (RATL) (\cite{chesson2003commonalities,onay2007intertemporal}). Our model accommodates both RSTL and RATL and shows that optimal information acquisition can differ dramatically under different time-risk preferences.

The optimal learning strategies that we identify are qualitatively similar to learning strategies that have been assumed in reduced form by many papers in the literature. For example, \cite{che2016optimal}, \cite{mayskaya2016dynamic}, and \cite{nikandrova2018dynamic} adopt a framework that restricts attention to Poisson signal processes in order to study optimal stopping with endogenous information. Poisson signals are also often assumed in the literature on strategic experimentation (see a survey by \cite{horner2016learning}). We show that Poisson learning has an optimization foundation under time-risk loving preferences.\footnote{To be clear, we do not show that \textit{all} Poisson learning strategies are optimal but rather that optimal strategies in our setting involve Poisson learning.} The Pure Accumulation strategy is also related to classic models on the timing of innovation introduced by \cite{dasgupta1980uncertainty} and \cite{lee1980market} (see a survey by \cite{reinganum1989timing}) which involve a deterministic time of innovation. The models in these papers assume a reduced-form learning process and are non-Bayesian. However, we show that the learning strategies in these papers can emerge endogenously in a Bayesian information acquisition  framework when agents have time risk-averse preferences. 

Our model also allows for Gaussian learning strategies. Gaussian signal processes are often assumed in reduced-form learning models (see for example \cite{moscarini2001optimal,ke2016optimal,liang2019dynamically,morris2017wald}). Also, \emph{drift-diffusion models} (DDM) of binary choice problems appear in \cite{ratcliff1998modeling} and \cite{fudenberg2015stochastic}. However, our results imply that Gaussian learning can not be justified by optimality except in the knife-edge case when agents have time-risk neutral preferences provided information can be acquired flexibly. 


The optimality of a greedy strategy is also the main result in \cite{liang2019dynamically}. However, the mechanisms in our papers are very different. \cite{liang2019dynamically}'s result crucially depends on the linear-Gaussian setup with exogenously given Gaussian information sources and holds for any time preferences. Our result allows for a flexible and endogenous choice of information sources, but crucially depends on time preferences. Also related is \cite{gossner2021attention} which studies a model with learning but with exogenous information sources and derives a greedy learning strategy that leads to the shortest stopping time (when the belief hits an exogenously given threshold) in the sense of first-order stochastic dominance.

 We model limits on the agent's learning resources via a constraint on the rate of entropy reduction. That is, the rate of resource depletion is determined by a \emph{uniformly posterior separable} (UPS) function. The rational inattention literature also typically models information costs or constraints using a UPS function (\cite{sims2003implications,matejka2014rational,steiner2016rational,NBERw23652}). Microfoundations for the UPS formulation can be found in \cite{frankel2018quantifying,NBERw23652,zhong2017linear,morris2017wald}. In our paper, the UPS information constraint ensures that the expected threshold hitting time is equalized for all exhaustive strategies, which allows us to isolate the role of time \emph{risk} in information acquisition. By Theorem 3 in \cite{zhong2017dynamic}, a UPS information constraint is both necessary and sufficient for the expected learning time to be equalized for all exhaustive strategies.

\section{Model}
This section presents a simple model of an agent who wants to learn over time about an unknown state. The unknown state $\omega$ takes values in $\{0,1\}$. At $t=0$, the agent believes that $\omega$ is 1 with probability $\mu\in (0,1)$. She receives a unit payoff when her posterior belief $\mu_t$ that $\omega$ is 1 reaches either an upper threshold $\overline{\mu}\in (\mu,1)$ or a lower threshold $\underline{\mu} \in (0,\mu)$. However, she is impatient and her utility is a decreasing function $\rho: \mathbb{R}_+ \rightarrow \mathbb{R}$ of the threshold-hitting time. When $\rho$ is convex, we say that the agent is \textit{time-risk loving}. When $\rho$ is concave we say that she is \emph{time-risk averse}. A common case when $\rho$ is convex arises when the agent discounts time at some exponential rate $r$: $\rho(t) = e^{-rt}$. On the other hand, a case when $\rho$ is concave is if the agent incurs a flow cost $c(\cdot)$ of delay that increases with time: $\rho(t) = 1- \int_0^tc(s)\,ds$.

The agent has great flexibility in how she can learn about $\omega$ but has limited resources and cannot learn infinitely fast. Let $\mathcal{M}$ denote the set of processes $\bm{\mu} = \{\mu_t; t\geq 0\}$ that take values in $[0,1]$ and satisfy a stochastic differential equation (allowing for jumps) of the form
\begin{equation} \label{belief processes}
	d\mu_t=  \sum_{i=1}^{N} (\nu^i(t,\mu_t)-\mu_t)\left[  dJ_t^i(\lambda^i(t,\mu_t))-\lambda^i(t,\mu_t) \,dt \right] + \sum_{j=1}^{M}\sigma^j(t,\mu_t) \,d Z_t^j
\end{equation}
with $\mu_0 = \mu$ for some positive integers $N$ and $M$ and functions $\{\nu^i\}_{i=1}^N$, $\{\lambda^i\}_{i=1}^N$, and $\{\sigma^j\}_{j=1}^M$.
Above, each $Z_t^j$ is a standard Brownian Motion and each $J_t^i$ is a Poisson point process that ticks at rate $\lambda^i(t,\mu_t^i)$. If $J_t^i$ ticks at time $t$,  the belief $\mu_t$ jumps to the point $\nu^i(t,\mu_t)$. The number of distinct points that the belief can jump to at time $t$ is the integer $N$ and the number of distinct Brownian Motions is the integer $M$. 

We assume that the agent can directly choose any belief process in $\mathcal{M}$ such that\footnote{Our restriction to jump-diffusion belief processes is without loss of generality within the larger class of c\'{a}dl\'{a}g processes such that \eqref{uncertainty reduction} is well defined. This follows from Theorem 1 in \cite{georgiadis2023information}.} 
\begin{equation}\label{uncertainty reduction}
\mathbb{E}\left[\frac{d}{dt}H(\mu_t)\Big|\mathcal{F}_t\right] \leq I
\end{equation} 
where $\{\mathcal{F}_t\}$ is the natural filtration of $\bm{\mu}$, $H$ is a $C^2$ 
strictly convex function defined on $[0,1]$, and $I\geq 0$ is a constant. $-H$ maps the agent's belief to its associated \emph{entropy}. Thus, equation \eqref{uncertainty reduction} is a constraint on the rate of entropy reduction. A special case is when $-H$ is Shannon's entropy so that \eqref{uncertainty reduction} amounts to a constraint on the well-known mutual information rate. Without loss of generality, we normalize
$H(\overline{\mu}) = H(\underline{\mu})=0$ and $I = 1$. This can be done by redefining $H$ to be $$\frac{1}{I}\left[H(\mu)-\frac{H(\overline{\mu})-H(\underline{\mu})}{\overline{\mu}-\underline{\mu}}(\mu-\underline{\mu})\right].$$
The same belief processes satisfy \eqref{uncertainty reduction} before and after this redefinition because beliefs are martingales (and thus, the drift of the second term in brackets above is zero). The normalization is convenient because, provided \eqref{uncertainty reduction} is binding at all times, the optional stopping theorem\footnote{See Theorem 3.22 of \cite{karatzas1998brownian}.} implies that the expected time remaining until a threshold is reached is simply the current entropy:
$$-H(\mu_t) = \mathbb{E}\left[H(\mu_{\tau}) - H(\mu_t)|\mathcal{F}_t\right] = \mathbb{E}\left[\tau-t|\mathcal{F}_t\right].$$

To state the agent's learning problem, let $\tau_{\bm{\mu}}$ be the first time that her beliefs reach a threshold:
$$\tau_{\bm{\mu}}: = \inf\{t| \mu_t \in  [0,\underline{\mu}]\cup [\overline{\mu},1] \}.$$  
Because $\tau_{\bm{\mu}}$ may be $\infty$ with positive probability for some belief processes (e.g., the agent can stop learning)
we set $\rho(\infty) = -\infty$ to ensure that the agent never selects these processes.\footnote{That is, we extend the domain and range of $\rho$.} 

She solves
 \begin{equation}\label{objective}
\max_{\bm{\mu} \in \mathcal{M}} \mathbb{E}[\rho(\tau_{\bm{\mu}})]
\end{equation}
such that \eqref{uncertainty reduction} holds.

Our simple model makes several assumptions in order to isolate the connection
between optimal learning and time-risk preferences which is the main focus of our paper. For example, we assume that the agent experiences no costs from learning (as long as \eqref{uncertainty reduction} is satisfied) and that she earns a common payoff regardless of the threshold that she ultimately hits. Though these assumptions are restrictive, we believe that our model aligns well with a number of economic applications as discussed in the Introduction. A key property of our model that allows for our main insights is that it accommodates general time-risk preferences. Indeed, our main contribution is to identify optimal learning strategies that apply across a range of time preferences and to highlight that it is the agent's attitude towards \textit{time risk} that determines the qualitative properties of these strategies. Common examples of time-risk loving preferences beyond exponential discounting include: (i) hyperbolic discounting where $\rho(t)=(1+\alpha t)^{-\gamma/\alpha}$ \citep{loewenstein1992anomalies}; (ii) generalized expected discounted utility: $\rho(t)=\phi(e^{-r t})$ where $\phi$ is increasing and convex 
\citep{dejarnette2020time}; and linear delay cost up until a deadline (iii) $\rho^T(t)=\max\{T-t,0\}$. Time-risk averse preferences are less commonly assumed in the literature though there is growing experimental evidence that agents may often be time-risk averse \citep{chesson2003commonalities,onay2007intertemporal}. As discussed at the start of this section, time-risk aversion can arise when agents have flow costs of delay that increase with time.

\section{Optimal Learning and Time-Risk Preferences}

In this section, we present our main results: a strategy that is optimal whenever the agent is time-risk loving and a strategy that is optimal whenever she is time-risk averse. These results illustrate the connection between optimal 
learning and time-risk preferences.

\subsection{Time-Risk Loving }
We first consider the case when the agent is time-risk loving. A formal statement of her optimal learning strategy is given below in Definition \ref{defi:greedy} but we begin by describing it informally here.  

An optimal strategy for the agent is 
\emph{Greedy Exploitation} and is illustrated in Figure \ref{greedy figure}. 
\medskip

\begin{figure}[h]
  \begin{center}
\include{Figures/Poisson}
\caption{Greedy Exploitation}
\label{greedy figure}
\bigskip
\begin{minipage}{0.9\linewidth}
 \footnotesize{\textit{Notes:} In dark green, we plot one possible realization of the belief path 
 $\bm{\mu}^{G}$ under the Greedy Exploitation strategy. In light green, we plot other possible realizations of  $\bm{\mu}^{G}$. The dashed lines with arrows represent jumps in the belief. The figure is computed for the case $H(\tilde{\mu})=\tilde{\mu}^2$.}
\end{minipage}
\end{center}
\end{figure}
Let
\begin{align*}
    d_{H}(\tilde{\mu},\hat{\mu})=H(\tilde{\mu}) - H(\hat{\mu}) - H'(\hat{\mu})(\tilde{\mu}- \hat{\mu})
\end{align*}
denote the Bregman divergence between any two beliefs $\hat{\mu}$ and $\tilde{\mu}$ under the function $H$.  In Figure \ref{greedy figure}, $\mu^*$ represents the unique belief that is equidistant in the Bregman divergence to the two thresholds: $d_H(\overline{\mu},\mu^*) = d_H(\underline{\mu},\mu^*)$. Initially, the agent's beliefs either jump to the threshold that is closer in Bregman divergence (in this case $\overline{\mu}$) or experience compensating drift towards the other threshold. By targeting the closer threshold, the agent greedily maximizes the ``chance" that her beliefs reach a threshold in the ``next instant." This is because the agent's beliefs can jump at a faster rate when she targets the closer threshold without violating her resource constraint \eqref{uncertainty reduction}. After some time, in the absence of a jump, her beliefs eventually drift to $\mu^*$. At this point, her beliefs may jump to either threshold. The jump rates to the respective thresholds are such that there is no net compensating drift, and so, in the absence of a jump, her beliefs remain stationary.

\begin{definition}\label{defi:greedy}
         The \textit{Greedy Exploitation} strategy    $\bm{\mu}^{G}$ is defined as follows. Let $\mu^* \in (0,1)$ be the unique belief such that 
    $d_H(\overline{\mu},\mu^*) = d_H( \underline{\mu},\mu^*)$,
    \begin{itemize}
        \item   While $\mu_t^{G}>\mu^*$, her beliefs evolve according to 
$$d\mu_t^{G} =  (\overline{\mu}-\mu_t^{G} )\left[  dJ_t^1(\lambda_t)-\lambda_t \,dt \right]$$
where $\lambda_t = I/d_H( \overline{\mu},\mu_t^{E} )$. 

\item While $\mu_t^{G}  = \mu^*$, her beliefs evolve according to
$$d\mu_t^{G}  = (\overline{\mu}-\mu_t^{G} )\,d J_t^2\left( \frac{\mu_t^{G} -\underline{\mu}}{\overline{\mu}-\underline{\mu}} \lambda^*\right)+(\underline{\mu}-\mu_t^{G} )\,d J_t^3\left( \frac{\overline{\mu}-\mu_t^{G} }{\overline{\mu}-\underline{\mu}}\lambda^*\right)$$
where $\lambda^* = 1/d_H( \overline{\mu},\mu^*)$. 

\item While $\mu_t^{G} <\mu^*$, her beliefs evolve according to 
$$d\mu_t^{G}  =  (\underline{\mu}-\mu_t^{G} )\left[  dJ_t^1(\lambda_t)-\lambda_t \,dt \right]$$
where $\lambda_t = 1/d_H(\underline{\mu},\mu_t^{G} )$. 
    \end{itemize}
   Above $J_t^1$, $J_t^2$, and $J_t^3$ are independent Poisson point processes with jump rates indicated in parentheses. 
\end{definition}

\begin{theorem}\label{risk loving optimal strategy}
If the agent is time-risk loving, then Greedy Exploitation is optimal.
\end{theorem}
Before we sketch the proof of Theorem \ref{risk loving optimal strategy}, we note that because Greedy Exploitation is uniformly optimal for all convex $\rho$, it must induce the riskiest distribution of threshold hitting times among all strategies that are exhaustive in that \eqref{uncertainty reduction} is satisfied at all points in time.  To make this precise, we first state the following definition.

\begin{definition}
    $\mathcal{T}=\left\{\tau_{\bm{\mu}}\big|\bm{\mu}\in\mathcal{M} \text{ such that \eqref{uncertainty reduction} binds at all $t$}\right\}$.
\end{definition}
The Greedy Exploitation strategy produces a threshold-hitting time that is maximal in the mean-preserving spread order among all threshold-hitting times in this set.
\begin{corollary} It holds that $\tau_{\bm{\mu}^{G}} \succeq_{\mathrm{mps}} \tau$ for each $\tau \in \mathcal{T}$.
\end{corollary}

This result hinges on our assumption that the constraint on learning is of the form in \eqref{uncertainty reduction}. Because of this assumption, all exhaustive strategies have the same expected threshold hitting time, which is equal to the initial entropy $-H(\mu)$.

\begin{proof}[Proof of Theorem \ref{risk loving optimal strategy}]
    The proof proceeds in seven steps. 
\bigskip

\textit{Step 1. Set of Basis Discount Functions}---We observe that any nonnegative convex $\rho$ can be written as a conical combination of functions of the form 
\begin{equation*}
\rho_T(t) = \max\{T-t,0\}
\end{equation*}
where $T\geq0$. Thus, if Greedy Exploitation is optimal for each $\rho_T$, $T\geq 0$, then it must be optimal for any nonnegative convex $\rho$. In fact, it will necessarily be optimal for \textit{any} convex $\rho$, including $\rho$ that may take on negative values. 
\begin{lemma}\label{Basis Functions}
    If Greedy Exploitation solves \eqref{objective} for each $\rho_T$ where $T\geq 0$, then it is optimal for any convex $\rho$.
\end{lemma}
\begin{proof}[Proof of Lemma \ref{Basis Functions}]
    See Theorem 3.6 in \cite{muller1996orderings}.\footnote{To apply Theorem 3.6 in \cite{muller1996orderings} recall that because $\rho$ is decreasing, any optimal strategy is exhaustive and thus has the expected threshold hitting time of $-H(\mu)$.}
\end{proof}

\textit{Step 2. Candidate Value Function and Some Notation}---Let $V(\mu,T) = \mathbb{E}\left[\rho_T(\tau_{\bm{\mu}^G})|\mu_0^G = \mu\right]$ denote the value function for Greedy Exploitation when $\rho = \rho_T$. That is, for each $T \geq 0$, let

\begin{equation}\label{candidate value function}
   V(\mu,T) = \begin{cases}
   \int_0^T (T-t) \lambda_t^G e^{-\int_0^t \lambda_z^G \, dz} \, dt,  &\mu \in (\underline{\mu}, \overline{\mu})\\
T, & \mu \in \{\underline{\mu}, \overline{\mu}\}.
\end{cases}
\end{equation}

In what follows, it is useful to observe that $\partial V(\mu,T)/\partial T = U(\mu,T)$ whenever $\mu \in (\underline{\mu}, \overline{\mu})$ where $$U(\mu,T) = \int_0^{T} \lambda_s^G e^{-\int_0^t \lambda_z^G \, dz} \, dt$$ 
is the probability that a jump arrives by time $t$ under Greedy Exploitation. It is easy to show that 
${\partial U(\mu,T)}/{\partial \mu}> 0 $ if $\mu \in [\mu^*,\overline{\mu})$ and  that ${\partial U(\mu,T)}/{\partial \mu}< 0 $ if $\mu \in (\underline{\mu}, \mu^*]$.

To ease the exposition, we adopt the following notation. We set $V_T(\mu)=V(\mu,T)$ and $U_T(\mu)=U(\mu,T)$. Also, given a function $f$ and any two beliefs $\nu$ and $\mu$, we let
 $$d_{f}(\nu,\mu) = f(\nu) - f(\mu) - f'(\mu)(\nu - \mu)$$
  whenever $f'(\mu)$ is well-defined. Note that if $f$ is convex, then $d_{f}$ is a Bregman divergence.
\bigskip

  \textit{Step 3. Verification Lemma}---To verify the optimality of Greedy Exploitation we use the following Lemma \ref{HJB lemma} which states that it suffices to prove that $V$ satisfies the Hamilton-Jacobi-Bellman (HJB) equation \eqref{HJB1}.

  \begin{lemma}\label{HJB lemma}
Given $T>0$, if $V$ in \eqref{candidate value function} satisfies 
\begin{equation}\label{HJB1}
 U_t(\mu) = \max \bigg{\{}\max_{\nu} \frac{d_{V_t}(\nu,\mu)}{d_H(\nu,\mu)}, \frac{{V_t''(\mu)}}{H''(\mu)}\bigg{\}}
\end{equation}
at each  $(\mu,t) \in (\underline{\mu},\overline{\mu}) \times [0,T]$ then $V_T(\mu)$ is equal to \eqref{objective} when $\rho = \rho^T$.
\end{lemma}
\begin{proof}[Proof of Lemma \ref{HJB lemma}]
    We first assert that condition \eqref{HJB1} is equivalent to 
\begin{equation}\label{HJB2}
\begin{aligned}
    U_t(\mu) =\max_{\{\nu^i\}, \{\lambda^i\} ,\sigma} \mathcal{A}^{\nu,\lambda, \sigma} V_t(\mu) \\
  \textrm{ s.t }  \mathcal{A}^{\nu,\lambda,\sigma} H(\mu_t)\leq 1
\end{aligned}
\end{equation}
 where   $\mathcal{A}^{\nu,\lambda,\sigma}$ is the operator defined for functions $f \in C^2 (\underline{\mu},\overline{\mu})$ by
	\begin{align*}
\mathcal{A}^{\nu,\lambda,\sigma}f(\mu)=\sum_i \lambda^i d_f(\nu^i,\mu)+\frac{1}{2}\sum_{j}(\sigma^j)^2 f''(\mu).
	\end{align*}
 That is, $\mathcal{A}^{\nu,\lambda,\sigma}$ is the infinitessimal generator for the compensated jump diffusion process \eqref{belief processes}. Because $\mathcal{A}^{\nu,\lambda,\sigma}$ is additively separable, it suffices to select a single jump point or volatility to achieve the max in \eqref{HJB2}. The jump point or volatility is chosen to maximize the ``bang-for-the-buck"---that is, the ratio of the drift of $V$ to the drift of $H$. Therefore, \eqref{HJB1} and \eqref{HJB2} must be equivalent.

Next, suppose that \eqref{HJB1} is satisfied. Consider an arbitrary admissible strategy
$\{\nu^i\}$, $\{\lambda^i\}$, $\{\sigma^j\}$ with induced first threshold-hitting time $\tau$. We have 
 \begin{align*}
	V_T(\mu) =& \mathbb{E}\Bigg[ V_{T-\tau\wedge T}(\mu_{\tau\wedge T})-\int_0^{\tau\wedge T}  \left[-U_{T-t}(\mu_t) + \mathcal{A}^{\nu,\lambda,\sigma} V_{T-s}(\mu_t)\right]\, d t \\ & + \sum_{j}\int_0^{\tau\wedge T} \frac{\partial  V_{T-t}(\mu_t)}{\partial \mu}\sigma_t^j \, d Z_t^j \\ &+  \sum_i \int_0^{\tau\wedge T} \left[V_{T-t}(\nu^i_t)-V_{T-t}(\mu_t) \right] \left(dJ_t^i(\lambda_t^i)-\lambda_t^i\,dt\right) \Bigg]\\ =&
\mathbb{E}\left[ V_{T-\tau\wedge T}(\mu_{\tau\wedge T})-\int_0^{\tau\wedge T}\left[-U_{T-t}(\mu_t) + \mathcal{A}^{\nu,\lambda,\sigma} V_{T-s}(\mu_t)\right]\, d t \right]\\
	\geq& \mathbb{E}\left[V_{T-{\tau \wedge T}}(\mu_{\tau \wedge T} )\right]\\
	\geq & \mathbb{E}\left[\rho^T(\tau)\right]
\end{align*}
where the first equality uses It\^{o}'s formula for jump diffusions and the fact that $\partial V/\partial T = U$ as noted in Step 2, the second equality follows from the fact that ${\partial  V_{T-t}(\mu_t)}/{\partial \mu}$ and $V_{T-t}$ are bounded which implies that the diffusion and jump terms are true martingales,\footnote{See Theorem 51 of \cite{protter2005stochastic}.} the first inequality follows from \eqref{HJB2}, and the last inequality follows from the definition of $V$. 
 \end{proof}

\textit{Step 4. $\{V_{T-t}(\mu_t^G)\}$ is a Martingale}---
The remaining steps verify that $V$ satisfies the conditions of Lemma \ref{HJB lemma}. We begin with Lemma \ref{second lemma} which states that if the inner and outer max on the right-hand side of \eqref{HJB1} is achieved by Greedy Exploitation then \eqref{HJB1} is statisfied. 
\begin{lemma}\label{second lemma}
At each $t \in [0,\infty)$ the following hold:
\begin{enumerate}
    \item If $\mu\geq\mu^*$, then 
    $$U_t(\mu) = \frac{d_{V_t}(\overline{\mu},\mu)}{d_{H}(\overline{\mu},\mu)}.$$
    \item    
    If $\mu \leq \mu^*$, then
 $$U_t(\mu) = \frac{d_{V_t}(\underline{\mu},\mu)}{d_{H}(\underline{\mu},\mu)}.$$
\end{enumerate}
\end{lemma}

\begin{proof}[Proof of Lemma 
\ref{second lemma}]
    Because $V_{T-t}(\mu_t^G) = \mathbb{E}\left[\rho^T(\tau_{\bm{\mu}^G})|\mu_t^G\right]$ and $\bm{\mu}^G$ is Markov it follows that $\{V_{T-t}(\mu_t^G)\}$ is a martingale for any given $T \geq 0$. By It\^{o}'s formula, the drift of $\{V_{T-t}(\mu_t^G)\}$ is zero if and only if conditions 1 and 2 of the lemma are satisfied. 
\end{proof}

\textit{Step 5. Unimprovable by Poisson Learning}---The following Lemma \ref{unimprovable Poisson} shows that Greedy Exploitation can not be improved on by any alternative Poisson learning strategy. 

 \begin{lemma}\label{unimprovable Poisson}
           At each $(\mu,t) \in (\underline{\mu},\overline{\mu}) \times [0,\infty)$ it holds that 
           \begin{equation}\label{Poisson eq}
           U_t(\mu) = \max_{\nu} \frac{d_{V_t}(\nu,\mu)}{d_H(\nu,\mu)}.
           \end{equation}
        \end{lemma}
           \begin{proof}[Proof of Lemma \ref{unimprovable Poisson}]
We will prove the lemma when $\mu>\mu^*$. The proof when $\mu\leq\mu^*$ is analogous. By Lemma \ref{second lemma}, it suffices to show that $\overline{\mu}$ achieves the max in \eqref{Poisson eq}.
We split the proof into three cases. 
\begin{itemize}[leftmargin=2em]
    \item Case 1: $\nu\geq \mu$. We will show that $\nu = \overline{\mu}$ is the global maximizer of ${d_{V_t}(\nu,\mu)}/{d_H(\nu,\mu)}$ in the region $\nu\geq\mu$. To start, we observe that
$$\frac{d}{d\nu}\frac{d_{V_t}(\nu,\mu)}{d_H(\nu,\mu)} = \frac{V_t'(\nu) - V_t'(\mu)}{d_H(\nu,\mu)} - \frac{d_{V_t}(\nu,\mu)}{d_H(\nu,\mu)^2}\left[H'(\nu) - H'(\mu)\right].$$
 This derivative is negative if and only if
\begin{equation}\label{prefoc}
\frac{V_t'(\nu) - V_t'(\mu)}{H'(\nu) - H'(\mu)} \geq \frac{d_{V_t}(\nu,\mu)}{d_H(\nu,\mu)}.
\end{equation}
which is equivalent to 
 \begin{equation}\label{foc}
 \frac{d_{V_t}(\overline{\mu},\mu) - d_{V_t}(\overline{\mu},\nu)}{d_{H}(\overline{\mu},\mu) - d_{H}(\overline{\mu},\nu)}\geq \frac{d_{V_t}(\nu,\mu)}{d_{H}(\nu,\mu)}.
 \end{equation}
Notice that \eqref{foc} holds with equality at $\nu = \overline{\mu}$. We will show that $\overline{\mu}$ and in fact any local extremum of ${d_{V_t}(\nu,\mu)}/{d_H(\nu,\mu)}$ in the region $\nu \in (\mu,\overline{\mu}]$ is a local maximum. This immediately implies that $\overline{\mu}$ must actually be a global maximum in this region.

At any local extremum \eqref{foc} holds with equality.  We can prove that all such local extrema are necessarily local maxima simply by proving that the derivative of the left-hand side of \eqref{foc} is negative. This is because the derivative of the right-hand side is always zero at a local extremum since the right-hand side expression is the objective ${d_{V_t}(\nu,\mu)}/{d_H(\nu,\mu)}$. 
The left-hand side of \eqref{foc} is decreasing in $\nu$ because
\begin{align*}
\frac{d}{d\nu} \frac{d_{V_t}(\overline{\mu},\mu) - d_{V_t}(\overline{\mu},\nu)}{d_{H}(\overline{\mu},\mu) - d_{H}(\overline{\mu},\nu)} &=\frac{d}{d\nu} \frac{U_t(\mu)d_{H}(\overline{\mu},\mu) - U_t(\nu)d_{H}(\overline{\mu},\nu)}{d_{H}(\overline{\mu},\mu) - d_{H}(\overline{\mu},\nu)} \\&< \frac{d}{d\nu} \frac{U_t(\mu)d_{H}(\overline{\mu},\mu) - U_t(\mu)d_{H}(\overline{\mu},\nu)}{d_{H}(\overline{\mu},\mu) - d_{H}(\overline{\mu},\nu)}\\&=0
\end{align*}
where we have used the fact that $d_{V_t}(\overline{\mu},\nu)/d_{H}(\overline{\mu},\nu) = U_t(\nu)$ from Lemma \ref{second lemma} and that $U_t(\nu)$ is increasing in $\nu$ as noted in Step 2.

\item Case 2: $\nu \in (\mu^*, \mu)$. In this region, (following the same steps as in Case 1) it is easy to show that
$d_{V_t}(\nu,\mu)/d_H(\nu,\mu)$ is nondecreasing if
\begin{equation}\label{foc flip}
 \frac{d_{V_t}(\overline{\mu},\mu) - d_{V_t}(\overline{\mu},\nu)}{d_{H}(\overline{\mu},\mu) - d_{H}(\overline{\mu},\nu)}\leq \frac{d_{V_t}(\nu,\mu)}{d_{H}(\nu,\mu)}.
\end{equation}
This is the same condition as \eqref{foc} except the inequality has flipped. 

As before, to determine whether a local extremum is a maximum or minimum it suffices to check how the left-hand side changes as $\nu$ increases. In this case, the left-hand side is increasing. This can be seen since 
\begin{align*}
\frac{d}{d\nu} \frac{d_{V_t}(\overline{\mu},\mu) - d_{V_t}(\overline{\mu},\nu)}{d_{H}(\overline{\mu},\mu) - d_{H}(\overline{\mu},\nu)}&= \frac{d}{d\nu} \frac{U_t(\mu)d_{H}(\overline{\mu},\mu) - U_t(\nu)d_{H}(\overline{\mu},\nu)}{d_{H}(\overline{\mu},\mu) - d_{H}(\overline{\mu},\nu)} \\&> \frac{d}{d\nu} \frac{U_t(\mu)d_{H}(\overline{\mu},\mu) - U_t(\mu)d_{H}(\overline{\mu},\nu)}{d_{H}(\overline{\mu},\mu) - d_{H}(\overline{\mu},\nu)}\\&=0\end{align*}
where we have used the fact that the denominator is negative in this case. Thus, in this region, any local extremum must be a local minimum. 
Thus, no point $\nu \in (\mu^*, \mu)$ can achieve the max in $\eqref{Poisson eq}$.
\item Case 3: $\nu \in [\underline{\mu},\mu^*]$.  Following analogous steps to those used to derive \eqref{foc}, we find that ${d_{V_t}(\nu,\mu)}/{d_H(\nu,\mu)}$ is decreasing in $\nu$ if and only if 
\begin{equation}\label{foctwo}
 \frac{d_{V_t}(\underline{\mu},\mu) - d_{V_t}(\underline{\mu},\nu)}{d_{H}(\underline{\mu},\mu) - d_{H}(\underline{\mu},\nu)} < \frac{d_{V_t}(\nu,\mu)}{d_{H}(\nu,\mu)}. 
 \end{equation}

We will prove that the left-hand side of \eqref{foctwo} is bounded above by $d_{V_t}(\overline{\mu},\mu)/{d_{H}(\overline{\mu},\mu)}$.
Thus, there can not be a point $\nu \in [\underline{\mu},\mu^*]$ that achieves a higher value than $d_{V_t}(\overline{\mu},\mu)/{d_{H}(\overline{\mu},\mu)}$, since if there was, at that point, $d_{V_t}(\nu,\mu)/{d_{H}(\nu,\mu)}$ would be decreasing in $\nu$. 

To show this, we first observe that 
\begin{equation}\label{lemma3eqn1}
 d_{V_t}(\underline{\mu},\mu) = d_{V_t}(\underline{\mu},\overline{\mu}) + d_{V_t}(\overline{\mu},\mu) - (\underline{\mu}-\overline{\mu})\left(V_t'(\mu) - V_t'(\overline{\mu})\right),
 \end{equation}
 and 
 \begin{equation}\label{lemma3eqn2}
  d_{H}(\underline{\mu},\mu) = d_{H}(\underline{\mu},\overline{\mu}) + d_{H}(\overline{\mu},\mu) - (\underline{\mu}-\overline{\mu})\left(H'(\mu) - H'(\overline{\mu})\right).
  \end{equation}
  Define $f(\mu)$ and $g(\mu)$ as
  \begin{equation}\label{lemma3eqn3}
  f(\mu) = d_{V_t}(\overline{\mu},\mu) - (\underline{\mu}-\overline{\mu})\left(V_t'(\mu) - V_t'(\overline{\mu})\right)
  \end{equation}
and 
\begin{equation}\label{lemma3eqn4}
g(\mu) = d_{H}(\overline{\mu},\mu) - (\underline{\mu}-\overline{\mu})\left(H'(\mu) - H'(\overline{\mu})\right).
\end{equation}
Since \eqref{prefoc} binds when $\nu = \overline{\mu}$, it follows that 
\begin{equation}\label{lemma3eqn5}
\frac{f(\mu)}{g(\mu)}=\frac{ d_{V_t}(\overline{\mu},\mu) - (\underline{\mu}-\overline{\mu})\left(V_t'(\mu) - V_t'(\overline{\mu})\right)}{ d_{H}(\overline{\mu},\mu) - (\underline{\mu}-\overline{\mu})\left(H'(\mu) - H'(\overline{\mu})\right)} = \frac{d_{V_t}(\overline{\mu},\mu)}{d_{H}(\overline{\mu},\mu)}.
\end{equation}
Also since $d_{V_t}(\overline{\mu},\mu^*)/{d_{H}(\overline{\mu},\mu^*)} = d_{V_t}(\underline{\mu},\mu^*)/{d_{H}(\underline{\mu},\mu^*)} $,
\begin{equation}\label{lemma3eqn6}
\frac{f(\mu^*)}{g(\mu^*)} = \frac{d_{V_t}(\underline{\mu},\overline{\mu}) + f(\mu^*)}{d_{H}(\underline{\mu},\overline{\mu})+ g(\mu^*)} \Rightarrow \frac{f(\mu^*)}{g(\mu^*)}=\frac{d_{V_t}(\underline{\mu},\overline{\mu})}{d_{H}(\underline{\mu},\overline{\mu})}.
\end{equation}

Thus, 
\begin{align*}
    \frac{d_{V_t}(\underline{\mu},\mu) - d_{V_t}(\underline{\mu},\nu)}{d_{H}(\underline{\mu},\mu) - d_{H}(\underline{\mu},\nu)}&=\frac{d_{V_t}(\underline{\mu},\overline{\mu}) + f(\mu) - d_{V_t}(\underline{\mu},\nu)}{d_{H}(\underline{\mu},\overline{\mu}) + g(\mu) - d_{H}(\underline{\mu},\nu)}\\ &=
    \frac{U_t(\mu^*) d_{H}(\underline{\mu},\overline{\mu}) + U_t(\mu)g(\mu) - U_t(\nu)d_{H}(\underline{\mu},\nu)}{d_{H}(\underline{\mu},\overline{\mu}) + g(\mu) - d_{H}(\underline{\mu},\nu)}\\
&\leq\frac{U_t(\mu^*) d_{H}(\underline{\mu},\overline{\mu}) + U_t(\mu)g(\mu) - U_t(\mu^*)d_{H}(\underline{\mu},\nu)}{d_{H}(\underline{\mu},\overline{\mu}) + g(\mu) - d_{H}(\underline{\mu},\nu)}\\
    &\leq U_t(\mu)\\ &= \frac{d_{V_t}(\overline{\mu},\mu)}{d_{H}(\overline{\mu},\mu)}.
\end{align*}
as desired. The first line uses \eqref{lemma3eqn1}, \eqref{lemma3eqn2}, \eqref{lemma3eqn3}, and \eqref{lemma3eqn4}. The second line uses \eqref{lemma3eqn5}, \eqref{lemma3eqn6}, and Lemma \ref{second lemma}. The third line uses the fact that $U_t(\nu)$ is decreasing for $\nu \in [\underline{\mu},\mu^*]$ as noted in Step 2. The last line uses Lemma \ref{second lemma}.
\end{itemize}
\end{proof}

\textit{Step 6. Unimprovable by Diffusion Learning}---The following Lemma \ref{brownian lemma} shows that Greedy Exploitation can not be improved on by diffusion learning. 
\begin{lemma}\label{brownian lemma}
       At each  $(\mu, t)\in(\underline{\mu},\overline{\mu}) \times [0,\infty)$, it holds that 
\begin{align*}
	U_t(\mu)\geq \frac{V_t''(\mu)}{H''(\mu)}.
\end{align*}
\end{lemma}

\begin{proof}
  
Recall from Step 2 the observation that $U_t'(\mu)>0$ when $\mu\in(\mu^*,\overline{\mu})$. Thus
\begin{multline*}
	U_t'(\mu)=\frac{d }{d\mu}\frac{d_{V_t}(\overline{\mu},\mu)}{d_H(\overline{\mu},\mu)} =\frac{-d_H(\overline{\mu},\mu)V''_{t}(\mu)(\overline{\mu}-\mu)+d_{V_t}(\overline{\mu},\mu)H''(\mu)(\overline{\mu}-\mu)}{d_H(\overline{\mu},\mu)^2}>0
\end{multline*}
which implies that
$$
	U_t(\mu)=\frac{d_{V_t}(\overline{\mu},\mu)}{d_{H}(\overline{\mu},\mu)}>\frac{V''_t(\mu)}{H''(\mu)}$$
 as desired. 
An analogous argument applies when $\mu\in(\underline{\mu},\mu^*)$. Weak inequality when $\mu=\mu^*$ follows from continuity.

\end{proof}
\textit{Step 7. Putting it All Together}---Lemmas \ref{unimprovable Poisson} and \ref{brownian lemma} imply that \eqref{HJB1} is satisfied by $V$ for any $T \geq 0$. Lemma \ref{HJB lemma} then implies that Greedy Exploitation is optimal for any discount function of the form $\rho_T$, $T\geq 0$. Lemma \ref{Basis Functions} then implies optimality for all convex $\rho$. The proof of Theorem \ref{risk loving optimal strategy} is complete.  

\end{proof}

\subsection{Time-Risk Averse}
When the agent is time-risk averse, her optimal learning strategy is \emph{Pure Accumulation}, illustrated graphically below in Figure \ref{fig:accu}.

\begin{figure}[h]
\begin{center}
\include{Figures/Accu}
\caption{Pure Accumulation}
\bigskip
\begin{minipage}{0.9\linewidth}
    \footnotesize{\textit{Notes.} The dark green curve represents one possible belief path of $\bm{\mu}^{P}$. The vertical
segments represent jumps. The light green curve represents another possible belief path of $\bm{\mu}^{P}$. The figure is computed for the case $H(\tilde{\mu})=\tilde{\mu}^2$.}
\label{fig:accu}
\end{minipage}
\end{center}
\end{figure}

As discussed in Section \ref{related literature}, the Pure Accumulation strategy is closely related to the \emph{suspense-maximal strategy} from \cite{ely2015suspense}.
Under this strategy, the agent's belief either jumps in the direction of the farther threshold or experiences compensating drift. When her belief jumps, it jumps to a point with the same entropy as her current belief so that all progress is made through drift. 

\begin{definition}\label{defi:accu}
     The \textit{Pure Accumulation} strategy is defined as follows. Let $\mu^H: [0,1]\setminus\{\mu^*\} \rightarrow [0,1]$ denote the function that maps a belief $\hat{\mu}$ to the unique belief $\mu^H(\hat{\mu}) \neq \hat{\mu}$ such that $H(\mu^H(\hat{\mu})) = H(\hat{\mu})$. Under Pure accumulation, the agent's beliefs evolve according to 
 $$d\mu_t^{P} = \left[\mu^H(\mu_t^{P}) - \mu_t^{P}\right]dJ_t(\lambda_t) - \lambda_t\left[\mu^H(\mu_t^{P}) - \mu_t^{P}\right]dt$$
        where
$J_t$ is a Poisson point process that ticks at rate $\lambda_t= 1/d_H(\mu^H(\mu_t^{A}), \mu_t^{A})$.
\end{definition}

\begin{theorem}
    If the agent is time-risk averse, then Pure Accumulation is optimal.
\end{theorem}
\begin{proof}
    Under Pure Accumulation, the agent is guaranteed to hit a threshold at the deterministic time $t = -H(\mu)$.
\end{proof}
Because Pure Accumulation entails no time risk, we immediately have the following result. 

\begin{corollary}
     It holds that $\tau \succeq_{\mathrm{mps}}\tau_{\bm{\mu}^{A}}$ for each $\tau \in \mathcal{T}$.
\end{corollary} 

\section{Concluding Discussion}\label{sec:conclude}
In this paper, we have studied the relationship between time-risk preferences and optimal information acquisition. We have shown that an optimal strategy for a time-risk loving agent is Greedy Exploitation. This strategy produces the riskiest distribution over threshold hitting times among all exhaustive strategies. On the other hand, an optimal strategy for a time-risk averse agent is Pure Accumulation. This strategy produces a deterministic threshold hitting time and thus entails no time risk. Both of these strategies are uniformly optimal up to the convexity or concavity of the utility function, provided the agent is impatient. Thus, they are immune to dynamic inconsistency.
In practice, agents may have time preferences that differ from the well-studied case of exponential discounting. Our analysis offers some insight into how these agents may seek to acquire information and the kinds of signal structures economists may consider using when modeling these agents. 

In order to illustrate the connection between learning and time-risk preferences as sharply as possible we have made a number of special assumptions. The assumptions of binary states, fixed stopping thresholds, and a capacity constraint on learning speed are critical because they ensure that all exhaustive strategies have the same expected threshold-hitting times.\footnote{Specifically, with a binary state and fixed thresholds, every learning strategy yields the same probability distribution over terminal beliefs by the martingale property which is not necessarily so with multiple states. That is, all learning strategies yield the same ``overall experiment." The capacity constraint then ensures that all learning strategies that result in the same  ``overall experiment" have the same expected threshold-hitting times.} The key reason why Greedy Exploitation and Pure Accumulation are optimal is because they respectively yield the maximal and minimal threshold-hitting times in the mean-preserving spread order among all exhaustive strategies. This allows us to emphasize that it is time \textit{risk} preferences that determines their optimality. The assumption that payoffs depend only on threshold hitting times and not on which threshold is hit allows us to derive closed-form solutions but is not critical for the economic insights.

Indeed, we anticipate that many of the qualitative properties of Greedy Exploitation or Pure Accumulation will persist under other model formulations, for example, with other costs of learning, multiple states, and endogenously chosen stopping thresholds (there are already examples in the literature where this is so as reviewed in Section \ref{related literature}). It is certainly possible to extend our model of optimal learning to accommodate these more general environments though explicit charaterizations will be rare.\footnote{Our solutions were based on a guess and verify approach that was possible because of the special structure of our setup. In more general setups guessing the optimal strategy seems more difficult.} The advantage of our special setup is that it is possible to solve for optimal strategies that are explicit and uniformly optimal for large classes of payoffs and, moreover, allows us to isolate the role of time risk for optimal learning.

There are two promising avenues to explore in future work. The first is to explore how our results may extend to the case when the agent is neither time-risk loving nor time-risk averse. For these more general preferences, what are the qualitative features of optimal information acquisition? A second avenue to explore is to try to embed our model of information acquisition into strategic settings where there are multiple agents in order to study the implications of flexible information acquisition in games.

\bibliographystyle{apalike2}
\bibliography{report.bib}
\end{document}

%% file: Figures/Poisson.tex
\tikzset{every picture/.style={line width=0.75pt}} 

\begin{tikzpicture}[x=0.75pt,y=0.75pt,yscale=-1,xscale=1]

\draw    (120,20.16) -- (124,20.16) ;
\draw    (120,179.92) -- (124,179.92) ;
\draw    (116,59.97) -- (120,59.97) ;
\draw    (116,100) -- (120,100) ;
\draw  [dash pattern={on 2.25pt off 1.5pt}]  (120,20.16) -- (450,20.16) ;
\draw  [dash pattern={on 2.25pt off 1.5pt}]  (120,179.92) -- (450,179.92) ;

\draw [color={rgb, 255:red, 126; green, 211; blue, 33 }  ,draw opacity=1 ][line width=1.5]    (120,59.97) .. controls (154.31,81.53) and (209.29,95.94) .. (272.73,100) ;
\draw [color={rgb, 255:red, 126; green, 211; blue, 33 }  ,draw opacity=1 ][line width=1.5]    (272.73,100) -- (450,100) ;
\draw [color={rgb, 255:red, 65; green, 117; blue, 5 }  ,draw opacity=1 ][line width=1.5]    (120,59.97) .. controls (146.39,75.17) and (168.45,84.5) .. (215.45,93) ;
\draw [color={rgb, 255:red, 65; green, 117; blue, 5 }  ,draw opacity=1 ][line width=1.5]  [dash pattern={on 1.5pt off 1.5pt}]  (215.45,92.89) -- (215.45,24.16) ;
\draw [shift={(215.45,20.16)}, rotate = 90] [fill={rgb, 255:red, 65; green, 117; blue, 5 }  ,fill opacity=1 ][line width=0.08]  [draw opacity=0] (6.97,-3.35) -- (0,0) -- (6.97,3.35) -- cycle    ;
\draw [color={rgb, 255:red, 126; green, 211; blue, 33 }  ,draw opacity=1 ][line width=1.5]  [dash pattern={on 1.5pt off 1.5pt}]  (323.64,100.24) -- (323.64,23.73) ;
\draw [shift={(323.64,19.73)}, rotate = 90] [fill={rgb, 255:red, 126; green, 211; blue, 33 }  ,fill opacity=1 ][line width=0.08]  [draw opacity=0] (6.97,-3.35) -- (0,0) -- (6.97,3.35) -- cycle    ;
\draw [color={rgb, 255:red, 126; green, 211; blue, 33 }  ,draw opacity=1 ][line width=1.5]  [dash pattern={on 1.5pt off 1.5pt}]  (323.64,100) -- (323.64,175.11) ;
\draw [shift={(323.64,179.11)}, rotate = 270] [fill={rgb, 255:red, 126; green, 211; blue, 33 }  ,fill opacity=1 ][line width=0.08]  [draw opacity=0] (6.97,-3.35) -- (0,0) -- (6.97,3.35) -- cycle    ;
\draw [color={rgb, 255:red, 126; green, 211; blue, 33 }  ,draw opacity=1 ][line width=1.5]  [dash pattern={on 1.5pt off 1.5pt}]  (158.18,78.16) -- (158.18,24.16) ;
\draw [shift={(158.18,20.16)}, rotate = 90] [fill={rgb, 255:red, 126; green, 211; blue, 33 }  ,fill opacity=1 ][line width=0.08]  [draw opacity=0] (6.97,-3.35) -- (0,0) -- (6.97,3.35) -- cycle    ;
\draw    (116,200) -- (450,200) ;
\draw    (120,205) -- (120,0) ;
\draw    (116,179.92) -- (120,179.92) ;
\draw    (116,20.16) -- (120,20.16) ;
\draw    (116,0) -- (120,0) ;

\draw (257,212.4) node [anchor=north west][inner sep=0.75pt]  [font=\small]  {Time\ $t$};
\draw (105,193.4) node [anchor=north west][inner sep=0.75pt]  [font=\small]  {$0$};
\draw (107,-5.6) node [anchor=north west][inner sep=0.75pt]  [font=\small]  {$1$};
\draw (100,12.4) node [anchor=north west][inner sep=0.75pt]  [font=\small]  {$\overline{\mu }$};
\draw (72.4,139) node [anchor=north west][inner sep=0.75pt]  [font=\small,rotate=-270]  {Belief\ $\mu _{t}$};
\draw (102,171.29) node [anchor=north west][inner sep=0.75pt]  [font=\small]  {$\underline{\mu }$};
\draw (100,51.64) node [anchor=north west][inner sep=0.75pt]  [font=\small]  {$\mu $};
\draw (100,89.4) node [anchor=north west][inner sep=0.75pt]  [font=\small]  {$\mu ^{*}$};
\draw (115,205.4) node [anchor=north west][inner sep=0.75pt]  [font=\small]  {$0$};

\end{tikzpicture}

\vspace{-3em}

%% file: Figures/Accu.tex
\tikzset{every picture/.style={line width=0.75pt}} 

\begin{tikzpicture}[x=0.75pt,y=0.75pt,yscale=-1,xscale=1]

\draw    (120,40.16) -- (124,40.16) ;
\draw    (120,199.92) -- (124,199.92) ;
\draw    (116,79.97) -- (120,79.97) ;
\draw  [dash pattern={on 2.25pt off 1.5pt}]  (120,40.16) -- (450,40.16) ;
\draw  [dash pattern={on 2.25pt off 1.5pt}]  (120,199.92) -- (450,199.92) ;
\draw    (116,220) -- (450,220) ;
\draw    (120,225) -- (120,20) ;
\draw    (116,199.92) -- (120,199.92) ;
\draw    (116,40.16) -- (120,40.16) ;
\draw [color={rgb, 255:red, 94; green, 74; blue, 74 }  ,draw opacity=0.26 ][line width=1.5]    (120,80) .. controls (215.68,51.27) and (314.68,47.27) .. (390,40) ;
\draw [color={rgb, 255:red, 94; green, 74; blue, 74 }  ,draw opacity=0.26 ][line width=1.5]    (120,160) .. controls (214.26,189.93) and (311.84,199.27) .. (390,200) ;
\draw [color={rgb, 255:red, 126; green, 211; blue, 33 }  ,draw opacity=1 ][line width=1.5]    (120,80) .. controls (157.42,67.93) and (216.63,57.27) .. (262.11,52) ;
\draw [color={rgb, 255:red, 126; green, 211; blue, 33 }  ,draw opacity=1 ][line width=1.5]    (262.11,191) -- (318.95,197) ;
\draw [color={rgb, 255:red, 126; green, 211; blue, 33 }  ,draw opacity=1 ][line width=1.5]    (318.95,46) -- (390,40) ;
\draw [color={rgb, 255:red, 126; green, 211; blue, 33 }  ,draw opacity=1 ][line width=1.5]  [dash pattern={on 1.5pt off 1.5pt}]  (262.11,52) -- (262.11,188) ;
\draw [shift={(262.11,192)}, rotate = 270] [fill={rgb, 255:red, 126; green, 211; blue, 33 }  ,fill opacity=1 ][line width=0.08]  [draw opacity=0] (6.97,-3.35) -- (0,0) -- (6.97,3.35) -- cycle    ;
\draw [color={rgb, 255:red, 126; green, 211; blue, 33 }  ,draw opacity=1 ][line width=1.5]  [dash pattern={on 1.5pt off 1.5pt}]  (318.95,198) -- (318.95,50) ;
\draw [shift={(318.95,46)}, rotate = 90] [fill={rgb, 255:red, 126; green, 211; blue, 33 }  ,fill opacity=1 ][line width=0.08]  [draw opacity=0] (6.97,-3.35) -- (0,0) -- (6.97,3.35) -- cycle    ;
\draw [color={rgb, 255:red, 65; green, 117; blue, 5 }  ,draw opacity=1 ][line width=1.5]    (120,80) -- (162.63,68) ;
\draw [color={rgb, 255:red, 65; green, 117; blue, 5 }  ,draw opacity=1 ][line width=1.5]    (162.63,172) -- (205.26,182) ;
\draw [color={rgb, 255:red, 65; green, 117; blue, 5 }  ,draw opacity=1 ][line width=1.5]    (205.26,60) -- (290.53,48) ;
\draw [color={rgb, 255:red, 65; green, 117; blue, 5 }  ,draw opacity=1 ][line width=1.5]    (290.53,195) -- (390,200) ;
\draw [color={rgb, 255:red, 65; green, 117; blue, 5 }  ,draw opacity=1 ][line width=1.5]  [dash pattern={on 1.5pt off 1.5pt}]  (162.63,68) -- (162.63,168) ;
\draw [shift={(162.63,172)}, rotate = 270] [fill={rgb, 255:red, 65; green, 117; blue, 5 }  ,fill opacity=1 ][line width=0.08]  [draw opacity=0] (6.97,-3.35) -- (0,0) -- (6.97,3.35) -- cycle    ;
\draw [color={rgb, 255:red, 65; green, 117; blue, 5 }  ,draw opacity=1 ][line width=1.5]  [dash pattern={on 1.5pt off 1.5pt}]  (205.26,180) -- (205.26,64) ;
\draw [shift={(205.26,60)}, rotate = 90] [fill={rgb, 255:red, 65; green, 117; blue, 5 }  ,fill opacity=1 ][line width=0.08]  [draw opacity=0] (6.97,-3.35) -- (0,0) -- (6.97,3.35) -- cycle    ;
\draw [color={rgb, 255:red, 65; green, 117; blue, 5 }  ,draw opacity=1 ][line width=1.5]  [dash pattern={on 1.5pt off 1.5pt}]  (290.53,48) -- (290.53,191) ;
\draw [shift={(290.53,195)}, rotate = 270] [fill={rgb, 255:red, 65; green, 117; blue, 5 }  ,fill opacity=1 ][line width=0.08]  [draw opacity=0] (6.97,-3.35) -- (0,0) -- (6.97,3.35) -- cycle    ;
\draw    (116,20) -- (120,20) ;

\draw (257,232.4) node [anchor=north west][inner sep=0.75pt]  [font=\small]  {Time\ $t$};
\draw (105,213.4) node [anchor=north west][inner sep=0.75pt]  [font=\small]  {$0$};
\draw (107,14.4) node [anchor=north west][inner sep=0.75pt]  [font=\small]  {$1$};
\draw (100,32.4) node [anchor=north west][inner sep=0.75pt]  [font=\small]  {$\overline{\mu }$};
\draw (72.4,159) node [anchor=north west][inner sep=0.75pt]  [font=\small,rotate=-270]  {Belief\ $\mu _{t}$};
\draw (102,191.29) node [anchor=north west][inner sep=0.75pt]  [font=\small]  {$\underline{\mu }$};
\draw (100,71.64) node [anchor=north west][inner sep=0.75pt]  [font=\small]  {$\mu $};
\draw (115,225.4) node [anchor=north west][inner sep=0.75pt]  [font=\small]  {$0$};

\end{tikzpicture}
\vspace{-3em}